\documentclass{article}

\PassOptionsToPackage{numbers, compress}{natbib}

\usepackage[final]{neurips_2022}

\usepackage[utf8]{inputenc} 
\usepackage[T1]{fontenc}    
\usepackage{hyperref}       
\usepackage{url}            
\usepackage{booktabs}       
\usepackage{amsfonts}       
\usepackage{nicefrac}       
\usepackage{microtype}      
\usepackage{xcolor}         
\usepackage{graphicx}

\title{Neural Network Prior Mean for Particle Accelerator Injector Tuning}

\author{%
  Connie Xu \\
  Duke University\\
  Durham, NC 27708 \\
  \texttt{connie.xu553@duke.edu} \\
\And
  Ryan~Roussel \\
  SLAC National Accelerator Laboratory\\
  Menlo Park, CA 94025 \\
  \texttt{rroussel@slac.stanford.edu} \\
  \And
  Auralee Edelen \\
  SLAC National Accelerator Laboratory\\
  Menlo Park, CA 94025 \\
  \texttt{edelen@slac.stanford.edu} \\
}

\begin{document}

\maketitle

\begin{abstract}
Bayesian optimization has been shown to be a powerful tool for solving black box problems during online accelerator optimization. The major advantage of Bayesian based optimization techniques is the ability to include prior information about the problem to speed up optimization, even if that information is not perfectly correlated with experimental measurements. In parallel, neural network surrogate system models of accelerator facilities are increasingly being made available, but at present they are not widely used in online optimization. In this work, we demonstrate the use of an approximate neural network surrogate model as a prior mean for Gaussian processes used in Bayesian optimization in a realistic setting. We show that the initial performance of Bayesian optimization is improved by using neural network surrogate models, even when surrogate models make erroneous predictions. Finally, we quantify requirements on surrogate prediction accuracy to achieve optimization performance when solving problems in high dimensional input spaces.
\end{abstract}

\section{Introduction}


Particle accelerators are important tools for scientific discovery in the fields of biology, chemistry and physics.
Optimization of accelerator control parameters, such as currents applied to magnetic elements or phases of radio-frequency cavities, during accelerator operation (i.e. ``online tuning'') is a tedious but often necessary part of any experimental facility's operation.
Due to their large number of components and the impact of variable external factors, such as temperature changes, accelerators must be continuously re-tuned and optimized to meet various beam quality objectives.
This requires hours of dedicated beamline time where teams of operational experts diagnose issues with beam quality and make corrections, for even small to medium size facilities.
This severely limits beam time that is available to experimenters (i.e. "users") thus reducing the facilities' overall scientific output.
With current technological advances in the fields of computer science and machine learning, it would be beneficial to have an automated or semi-automated algorithm take care of normal beamline tuning, reducing downtime while also allowing human experts to tackle more challenging operational and design problems.

Bayesian optimization \cite{shahriari_taking_2016, greenhill_bayesian_2020} provides a framework for global optimization, while significantly reducing the number of physical observations needed to find solutions while also taking into account functional noise.
In this method, physical observations are combined with a kernel, which describes the overall functional behaviour, to create what is commonly referred to as a Gaussian Process (GP) model, which predicts the value and uncertainty of a target function \cite{rasmussen_gaussian_2006}.
Using this prediction, an optimizer can then choose input points that are likely to be ideal, before a physical measurement is made.
This method has been successfully used to efficiently optimize single objective problems at accelerators such as the Linac Coherent Light Source (LCLS) and the Stanford Positron Electron Asymmetric Ring (SPEAR3), with a lower number of observations needed than Nelder-Mead Simplex and RCDS algorithms \cite{hanuka_online_2019, mcintire_bayesian_2016, duris_bayesian_2019, kirschner_adaptive_2019}.

Despite this progress, optimizing accelerators remains a challenge due to the large number of tunable parameters and the so-called `curse of dimensionality`. 
Increases in input space dimensionality generally requires a large number of observations to find extrema.
This presents a problem for using Bayesian optimization due to the computational complexity scaling of Gaussian processes.
It is desirable to incorporate prior physics information into the Gaussian process model in order to improve predictive accuracy with fewer observations, which corresponds to improvements in optimization sample efficiency.

An initial attempt at incorporating physics information used specialized kernel functions to express function value correlations between input variables via a Hessian matrix, calculated from simulations \cite{hanuka_physics_2021}.
Alternatively, the Gaussian Process's prior mean function, which predicts the function value in the absence of experimental observations, can be used to incorporate prior physics-based knowledge about the target function \cite{hwang:ipac2022-tupost053, https://doi.org/10.48550/arxiv.2211.06400}.
To be effective in the context of Bayesian optimization, the prior mean function should be fast-executing and differentiable (to allow for gradient-based optimization of the acquisition function).
We propose to use neural network surrogate models as prior mean functions. 
Surrogate models have been trained to predict beam properties along the accelerator using data collected from physics simulations, and are increasingly available for accelerators \cite{edelen_machine_2020}.
However, surrogate models based on simulation data do not perfectly replicate experimental measurements due to imperfections in the simulation, which raises the question of how accurate the prior mean from the surrogate model would need to be in order to provide a performance improvement for Bayesian optimization.
This work investigates how accurate prior mean surrogate models must be in order to improve the convergence rate of Bayesian optimization in the context of accelerator optimization.

\section{Methods}
\begin{figure}[h]
    \includegraphics[width=\linewidth]{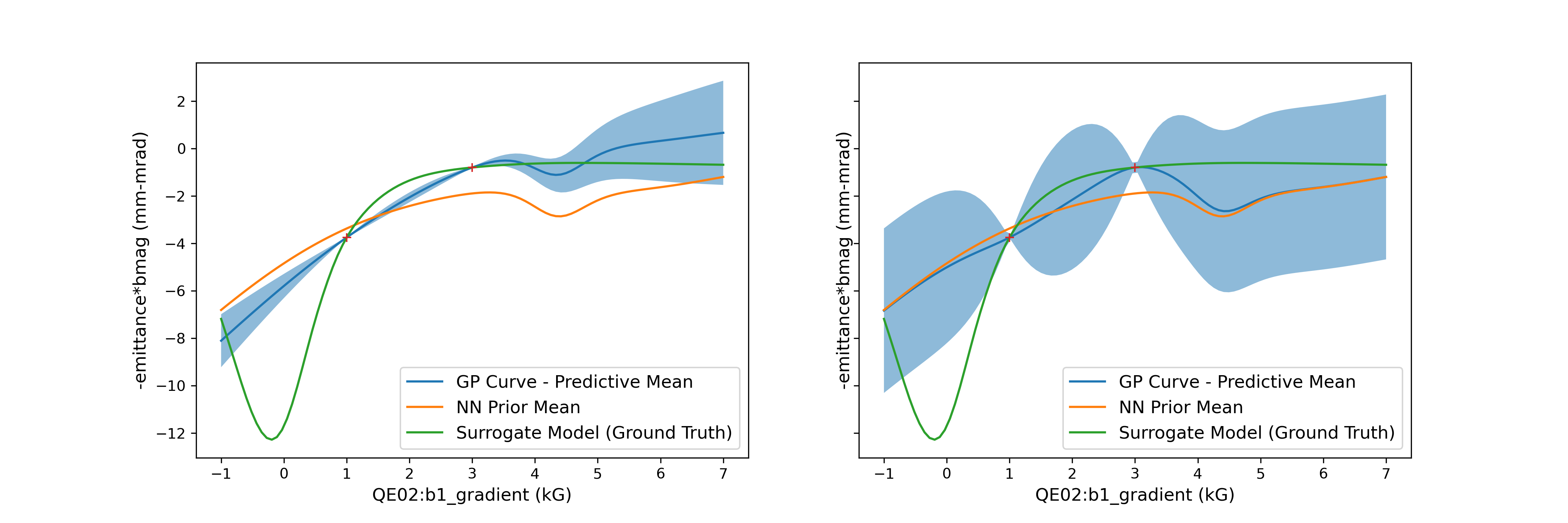}
    \caption{Sample fittings of a GP with an expressive neural network prior to 2 ground truth points in the input space. (Left) shows the original GP predictive mean after fitting, while (right) uses a smaller lengthscale to better demonstrate how the predicted mean curve returns to the prior mean in areas with no data samples. The predictive mean is used to find the next input to Bayesian Optimization, so an expressive prior that already resembles the ground truth will make it faster for the acquisition function to find a better objective value.}
   \label{fig:gp_prior_comparison}
\end{figure}

Bayesian optimization has been used at LCLS to find accelerator control parameters that minimize a product of the beam "emittance" $\varepsilon$ and beam "mismatch" $b_{mag}$. 
We prototype this optimization approach using a surrogate model of the LCLS accelerator injector. 
The objective is a function of 9 input parameters controlling the strength of magnetic focusing elements in the injector.
 
We modify previous Bayesian optimization algorithms used at LCLS to include a custom prior mean. 
We train neural network models on simulated samples collected from the LCLS injector. 
This model is specified as the mean function of the Gaussian Process and will be used to model the initial behavior of the objective function. 
The Gaussian Process is used to estimate the objective after every iteration of Bayesian optimization. 
The system gives data samples to the GP, and the GP uses these samples to generate the predictive mean function from the prior mean and covariance function. The acquisition function that we use, Upper Confidence Bound (UCB), finds the inputs most likely to yield an optimum and returns it to the system. 
The system finds the true value of the input and adds the new data sample to its samples, and the process repeats for a specified number of iterations. 
We used PyTorch \cite{paszke_pytorch_2019} for all models and Bayesian optimization implementations to aid end-to-end differentiability, and we used BoTorch \cite{balandat_botorch_2020} and GPyTorch libraries to implement the Bayesian optimization. 

\begin{figure}[h]
    \includegraphics[width=\linewidth]{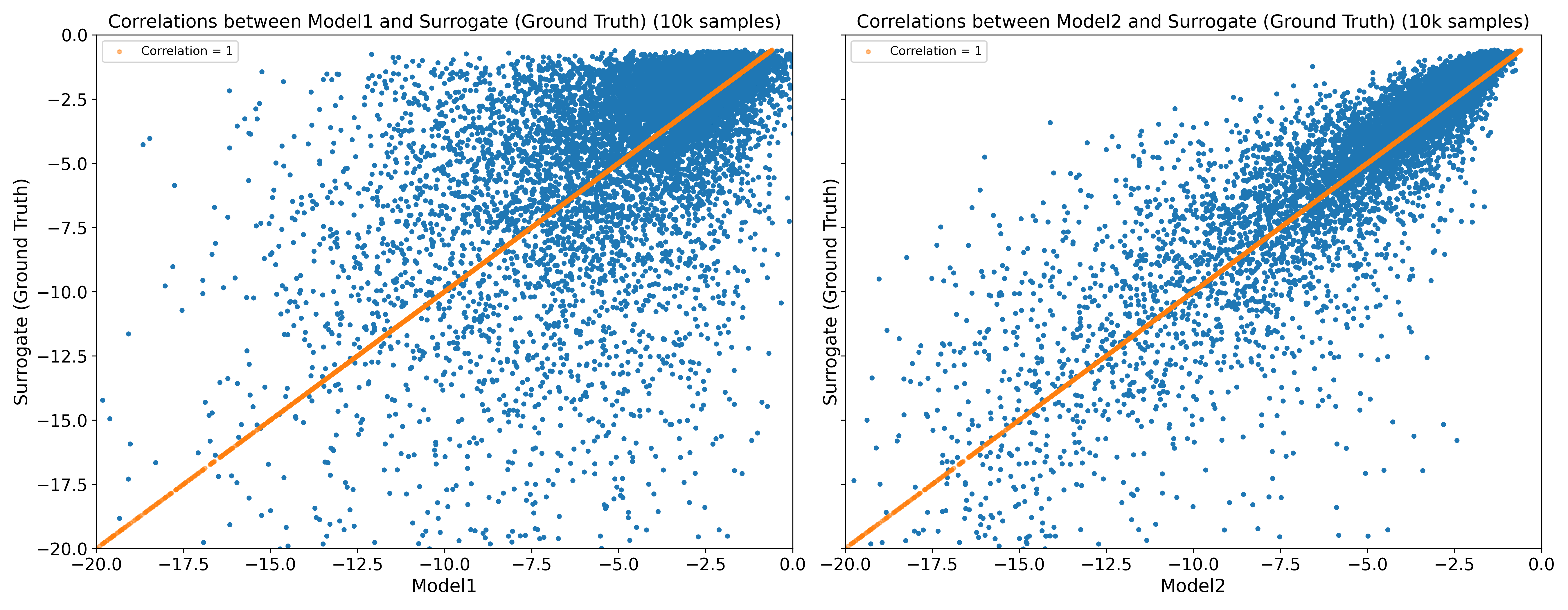}
    \caption{Correlation plots and values between each neural network, (left) Model1 and (right) Model2, prior mean and the ground truth. The plotted points and correlation values are calculated from the points out of 10k random data samples in between -20 and 0. Model2 has a higher positive correlation with the ground truth than Model1 does, which suggests that Model2 is more accurate to the ground truth and should have better Bayesian Optimization performance.}
   \label{fig:correlation}
\end{figure}

We investigated the effect of model accuracy on optimization performance by training two neural nets of different complexities on meshgrids of data samples generated from the injector model. The first neural network has a simpler structure (fewer hidden layers and nodes per layer) and is trained on a meshgrid of $3^9$ data samples; in contrast, the second neural network has a more complex structure and is trained on a meshgrid of $4^9$ data samples. We removed any data samples where the objective was less than -20 to eliminate large negative values that were disrupting model accuracy. To compare how well the neural networks reflect the ground truth, we performed parameter scans across the 9 input variables and measured the correlations between each neural net and the ground truth surrogate model, shown in Fig.~\ref{fig:correlation}. Results suggested that the more complex model seems to follow the ground truth behavior more closely and is thus the higher accuracy model.  

\begin{figure}[h]
    \includegraphics[width=\linewidth]{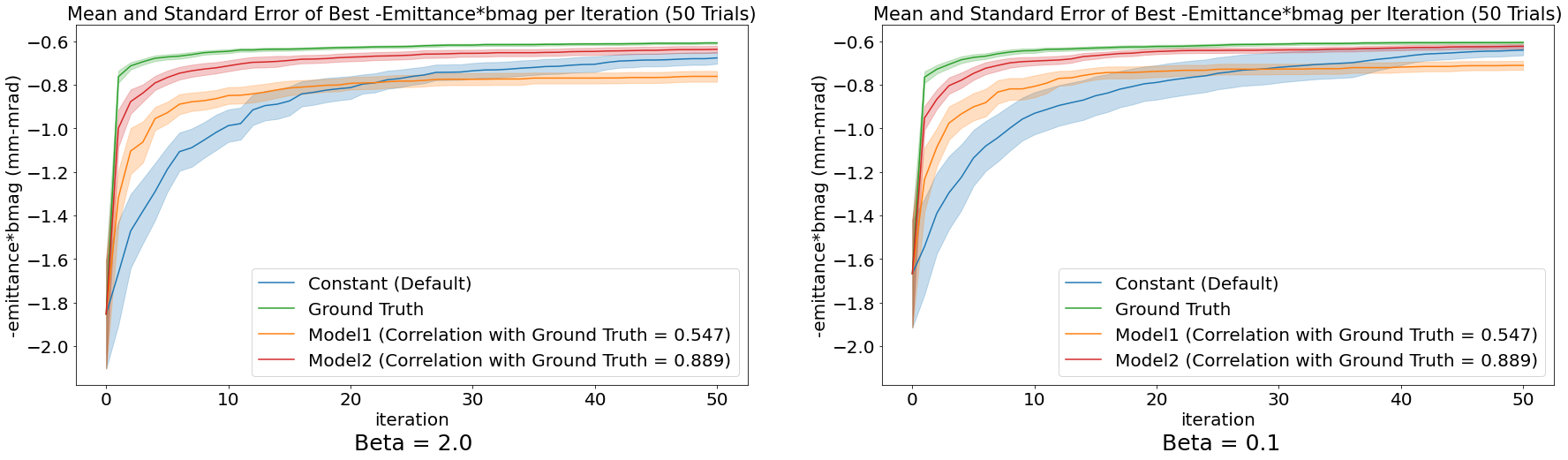}
    \centering
    \caption{Bayesian Optimization performance comparisons for beta=2.0 and beta=0.1 between the constant prior mean, the ground truth prior mean, and the less accurate (Model1) and more accurate (Model2) neural network prior means. Shading denotes one standard deviation uncertainty in best optimization performance.}
   \label{fig:combined_result}
\end{figure}

After training the two neural networks, we incorporated the models into custom prior mean classes, which made it possible for us to specify them as prior means for Gaussian Processes that we initialize for Bayesian optimization. 

We ran Bayesian optimization comparison trials for four priors: a constant mean that is retrained at every optimization step, a fixed injector surrogate mean, and fixed less accurate and more accurate neural nets, denoted as Model1 and Model2 respectively. We tested performance for Bayesian optimization with a using the Upper Confidence Bound \cite{snoek_practical_2012} acquisition function with $\beta=2.0$ and $\beta=0.1$, retraining the kernel hyperparameters at each step. For each experiment, we ran 50 trials, starting with 3 randomly initialized data samples given by the injector model and tracked the maximum objective value found per iteration over a total of 50 iterations. 

\section{Results}
We compare the Bayesian optimization performances of the two neural nets against the constant mean and the injector surrogate mean ("ground truth") in Figure \ref{fig:combined_result}. Model1 surpasses the constant prior in earlier iterations of Bayesian Optimization but stagnates around the 30th iteration, allowing the constant prior to eventually perform better after 50 iterations.  Meanwhile, Model2 consistently outperforms both the constant prior and Model1 after 50 iterations. Its performance is similar to that of the injector surrogate mean, and it converges to a value that is close to the one reached by the injector surrogate mean.
During the later stages of tuning we believe that greater inaccuracies with modeling the ground truth is causing Model1's inability to find better optimum and its worse performance compared to the constant prior, since it is not retrained during optimization.
A possible strategy for overcoming this challenge is to specify an extra network layer that is trainable during optimization, but has priors specified over the training weights.

\section{Conclusions}
Fast-executing neural network surrogate models of accelerator systems are increasingly available in the accelerator community, but at present these are not widely used in machine tuning compared with techniques like Bayesian optimization with Gaussian Processes. In this work we have investigated the use of neural networks as a prior mean function specification for use in Bayesian optimization based tuning of accelerator parameters.
We found that if the prior mean function predictions show reasonable correlation with experimental measurements then their use can improve initial optimization performance.
However, once a significant amount of data has been collected during optimization, refitting the prior mean improves optimization performance relative to fixed prior mean functions.
Combining the advantages gained by both cases is a topic of future study.
The code used to conduct this study can be found in the Github repository \url{https://github.com/sylvia5monthes/BO_Expressive_Priors_Comparisons}.

\section{Impact Statement}
This work is expected to have impact on the performance of accelerator optimization algorithms in the future. As the work is limited within the scope of accelerator science we expect there to be no ethical aspects. The main societal impacts are the improvement of accelerator operations and their end applications in science, industry, and medicine. For example, accelerators are used in medical applications (e.g. cancer radiation therapy), and fine-tuning of beams in those cases can enable higher-precision treatment.

\bibliographystyle{unsrtnat}
\bibliography{my_bib}
\setcitestyle{numbers,open={[},close={]},citesep={,}}

\end{document}